# Coupling and competition between ferroelectricity, magnetism, strain and oxygen vacancies in $A$MnO$_3$ perovskites

Astrid Marthinsen[2*], Carina Faber[1*], Ulrich Aschauer[1,3], Nicola A. Spaldin[1], Sverre M. Selbach[2]

[1]*Materials Theory, ETH Zürich, Wolfgang-Pauli Strasse 27, CH-8093 Zürich, Switzerland*

[2]*Department of Materials Science and Engineering, Norwegian University of Science and Technology, NO-7491 Trondheim, Norway*

[3]*Department of Chemistry and Biochemistry, University of Bern, Freiestrasse 3, CH-3012 Bern, Switzerland*

**\*Equal contribution**

## Abstract

We use first-principles calculations based on density functional theory to investigate the interplay between oxygen vacancies, $A$-site cation size / tolerance factor, epitaxial strain, ferroelectricity and magnetism in the perovskite manganite series, $A$MnO$_3$ ($A$=Ca$^{2+}$, Sr$^{2+}$, Ba$^{2+}$). We find that, as expected, increasing the volume through either chemical pressure or tensile strain generally lowers the formation energy of neutral oxygen vacancies consistent with their established tendency to expand the lattice. Increased volume also favors polar distortions, both because competing rotations of the oxygen octahedra are suppressed and because coulomb repulsion associated with cation off-centering is reduced. Interestingly, the presence of ferroelectric polarization favors ferromagnetic over antiferromagnetic ordering due to suppressed antiferromagnetic superexchange as the polar distortion bends the Mn-O-Mn bond angles away from the optimal 180$^\text{o}$. Intriguingly, we find that polar distortions compete with the formation of oxygen vacancies, which have a higher formation energy in the polar phases; conversely the presence of oxygen vacancies suppresses the onset of polarization. In contrast, oxygen vacancy formation energies are lower for ferromagnetic than antiferromagnetic orderings of the same structure type. Our findings suggest a rich and complex phase diagram, in which defect chemistry, polarization, structure and magnetism can be modified using chemical potential, strain, and electric or magnetic fields.

# Introduction

Strongly correlated oxides exhibit a vast range of functional properties owing to their intimate coupling between electronic, magnetic and structural degrees of freedom[1, 2]. The energetics of the various interactions tend to be similar in magnitude, and thus even small external perturbations, such as strain imposed on thin films through coherent heteroepitaxy, can have a large influence on the functionality[3]. An example is the generation of ferroelectric polarization in otherwise non-polar magnetic perovskite oxides such as $EuTiO_3$[4] and $SrMnO_3$[5], circumventing the usual contraindication between magnetism and ferroelectricity[6]. In both cases first-principles calculations based on density functional theory were instrumental in first predicting then explaining the observed behavior[7, 8].

In addition to these intrinsic behaviors, point defects such as oxygen vacancies and cation non-stoichiometry, which are always present in perovskite oxides, may significantly alter their properties or even enable new functionalities[9, 10]. For example, oxygen non-stoichiometry often introduces carriers that can lead to insulator-metal transitions[11] or superconductivity[12], or screen polar discontinuities to stabilize unusual domain structures[13, 14]. And of course in devices based on ionic transport, such as solid oxide fuel cells, point defects are crucial for the functionality[15, 16]. Based on the well-established coupling between defect chemistry and volume in bulk ceramics[17], it has recently been recognized that point defects should also couple with strain, which could therefore be used as a parameter for tuning or controlling the defect concentration in oxide thin films[10]. Again, first-principles calculations based on density functional theory have been invaluable in predicting and interpreting this behavior[18].

In this work we use first-principles calculations based on density functional theory to study the interplay between strain, ferroelectricity, oxygen vacancy formation, composition and magnetism in transition-metal oxides. We choose as our model system the series of II-IV perovskite-structure manganites, $CaMnO_3$, $SrMnO_3$ and $BaMnO_3$. Our motivation is two-fold: First the size of the divalent $A$-cations increases monotonically from $Ca^{2+}$ to $Sr^{2+}$ to $Ba^{2+}$ allowing for a systematic study of so-called chemical pressure effects. Second, a number of excellent computational and experimental studies for individual members of the series already exist; we gather the earlier findings here and complement them with our own new calculations to provide a complete picture. We begin by reviewing the behavior of $CaMnO_3$, for which the strain dependence of both the intrinsic structure [19, 20] and the defect formation energies[18] have been previously studied. CaMnO$_3$ maintains the a$^-$a$^-$c$^+$ tilt system of its bulk orthorhombic perovskite structure [21, 22] over the whole range of accessible strains, lowering its symmetry from *Pbnm* to *Pmc*2$_1$ when it becomes ferroelectric at a predicted tensile strain value of a few per cent; at the largest tensile strains achieved experimentally to date (2.1%) incipient ferroelectricity has been observed[20]. Regarding its defect chemistry, the oxygen vacancy formation energy is predicted to decrease with tensile strain, with vacancy sites in Mn-O-Mn bonds in the epitaxial plane favored. We then extend our theoretical study to SrMnO$_3$, which exists as a bulk compound both in the 4H hexagonal structure[23, 24] and the competing perovskite phase[25-28]. The perovskite phase is close to the ideal cubic structure with small octahedral tilts and rotations found computationally[8] but not yet observed experimentally. Small tensile or compressive strain was predicted[8] and subsequently confirmed[5] to induce ferroelectricity in the perovskite variant, with unusual transport behavior found in the experiments suggesting an intriguing interplay between strain, domain structure and oxygen vacancy formation. Finally, we discuss the case of BaMnO$_3$, for which the ground state is the non-perovskite hexagonal 2H structure[29] and for which the hypothetical metastable bulk perovskite structure lacks octahedral tilts or rotations and is already polar even without additional strain[30]. The trend of reduced octahedral rotations down the series, as well as increased tendency to ferroelectric polarization are consistent with the increasing *A*-site cation radius and resulting increasing tolerance factor from Ca$^{2+}$ to Sr$^{2+}$ to Ba$^{2+}$.

Our consolidation of literature data and additional calculations confirm previously observed or suggested relationships between polarization and strain and between oxygen vacancy formation energy and strain, and extend them to consider the interplay with chemical pressure from the *A*-site cation radius. Our main new finding is that oxygen vacancies *compete* with ferroelectric distortions as a response of the material to biaxial strain. As a result, the attempt to induce ferroelectric polarization using tensile epitaxial strain can be stymied by the formation of oxygen vacancies, which increase the effective lattice constant of the material and hence lower the strain transmitted to the film. Conversely, our calculations suggest that the onset of polar distortions destabilizes oxygen vacancies. In addition, while we do not provide an exhaustive study of the magnetism in the series, we find an intriguing correlation between the onset of ferroelectric polarization and ferromagnetism, the details of which will be a subject of future work.

The remainder of this paper is organized as follows: In the next section we provide details of the computations that are performed in this work. Then in the first results section we describe the effect of biaxial strain on the crystal structure, lattice dynamics and magnetic order in the stoichiometric compounds, paying particular attention to the interplay between the structural, electronic and magnetic degrees of freedom. The second results section adds oxygen vacancies and chemical potential to the list of coupled and competing variables, and we calculate how the oxygen vacancy formation is influenced by biaxial strain, as well as ferroelectric polarization and magnetism. Finally, we provide a summary of the coupled and competing variables that influence thin films of the II-IV manganites, and discuss the implications of our results.

# Computational details

Density functional theory (DFT) calculations were performed using the VASP package (version 5.3.3)[31, 32] within the generalized gradient approximation (GGA)[33]. We used the PBEsol exchange-correlation functional since it is known to accurately reproduce the experimental lattice constants, to which polar transitions and magnetic orderings are highly sensitive. The electronic wave functions were expanded using the projector augmented wave (PAW) method[34] with valence electron configurations of $4s^2$ for Ca, $4s^24p^65s^2$ for Sr, $5s^25p^66s^2$ for Ba, $3p^64s^23d^5$ for Mn and $2s^22p^4$ for O and a plane wave basis set energy cutoff of 650 eV. We imposed an effective Hubbard U correction[35] of 3eV on the Mn 3d electrons in all cases; for $CaMnO_3$, this value has been shown to correctly reproduce the experimentally determined density of states[36] and the ground-state G-type antiferromagnetic order[37]. All calculations were performed on 40-atom 2x2x2 supercells of the 5-atom primitive cubic cell, with a $\Gamma$-centered 4×4×4 Monkhorst-Pack k-point mesh[38] to sample the Brillouin zone; we repeated the calculations for $SrMnO_3$ using a 6x6x6 k-point grid which changed the oxygen vacancy formation energies by only 0.01 eV. We studied both G-type antiferromagnetic (AFM) and ferromagnetic (FM) orderings for all three materials.

Homoepitaxial biaxial strain was simulated as described in detail in Ref. 3 by fixing the in-plane *a* and *b* lattice parameters (in the *Pbnm* setting) to be equal in length and at 90° to each other and letting the out-of-plane *c* lattice parameter and internal coordinates relax to their lowest energy positions. In this orientation the rotations of neighboring octahedra are out-of-phase in the strain plane and in-phase in the perpendicular direction; the resulting strained thin-film structures are pseudo-orthorhombic. To test the effect of axis orientation, we also performed calculations for $SrMnO_3$ with *b*-axis out-of-plane orientation, and obtained similar trends for the oxygen vacancy formation energies with absolute values differing by ~0.01eV. We defined the 0% value of strain to correspond to the structure with the same in-plane surface area as the relaxed G-type AFM bulk structure for both AFM and FM magnetically ordered strained films. Note that this structure is higher in energy than the fully relaxed ground state as a result of setting the in-plane lattice parameters to be equal and at 90° to each other, and, in the FM case, from taking the AFM values for their length. The ionic positions and the out-of-plane axis (for both FM and AFM orderings) were relaxed using the conjugate-gradient algorithm until Hellmann-Feynman forces on all ions were below 1 meV Å.

Lattice instabilities were calculated at the R, Γ, M- and X-points of 2x2x2 supercells of the 5-atom ideal *Pm*-3*m* perovskite primitive cell using the frozen phonon approach[39] and analyzed using the phonopy interface[40]. The Berry-phase formalism was used to evaluate the electronic contribution to the ferroelectric polarization[41, 42].

Oxygen vacancy formation energies $\Delta E_{\text{form}}$ were computed from:

$$\Delta E_{\text{form}}(\epsilon, \mu_O) = E_{\text{tot},V_O} - E_{\text{tot,stoich}} + \mu_O,$$

where $\epsilon$ is the imposed biaxial strain, $E_{\text{tot},V_O}$ is the total energy of the 40-atom supercell containing an oxygen vacancy, and $E_{\text{tot,stoich}}$ is the total energy of the stoichiometric 40-atom supercell. For structural optimization of supercells containing oxygen vacancies, the lattice parameters were fixed to those of the stoichiometric case. Additional calculations for $CaMnO_3$ and $SrMnO_3$ in which the out-of-plane lattice parameter was re-relaxed gave the same trends for oxygen vacancy formation energies with absolute values differing by ~0.05eV.

# Results and Discussion

**Stoichiometric $A$MnO$_3$: ground state structures and strain**

We begin by reviewing the structure and properties of the stoichiometric perovskite compounds without oxygen vacancies.

Crystal and Magnetic Structure

Our calculated ground state structures for all three perovskite materials are shown in Fig. 2. Consistent with the literature, we obtain the orthorhombic *Pbnm* structure with large rotations of the oxygen octahedra and G-type AFM ordering of the Mn$^{4+}$ moments as the ground state for CaMnO$_3$. For G-type AFM BaMnO$_3$ we obtain a polar structure of *Amm*2 symmetry with polarization along the [110] direction and without tilts or rotations of the oxygen octahedra, again consistent with an earlier report[30]. Within our choice of parameters, the *P4mm* variant with polarization along [001] and FM ordering is slightly (3.8 meV per formula unit) lower in energy. In much of the following we analyze the AFM case to facilitate comparison with the other members of the series, as well as because the FM ordering leads to a closing of the DFT gap and so its Berry phase polarization cannot be defined. Reflecting its intermediate $A$-cation size, the structure of SrMnO$_3$ is between CaMnO$_3$ and BaMnO$_3$, with no polarization in its ground state and very small octahedral rotations; the *Pbnm* and $R\bar{3}c$ space groups are indistinguishable in energy (less than 0.5 meV per formula unit difference) within the accuracy of our calculations. Its ground-state magnetic ordering is G-type AFM. In Table 1 we summarize our calculated lattice constants for the three materials in their fully relaxed perovskite structures, compared to experiment and earlier calculations using different exchange-correlation functionals. Our values show good agreement with measured values where available, as well as with earlier calculations using the local density approximation (LDA) and GGA PBEsol functionals.

This evolution of the ground-state crystal structures with A-site cation size is reflected in the calculated phonon spectra for the high-symmetry cubic G-AFM reference structures (Fig. 3). Negative frequencies in the figure correspond to imaginary frequency phonons, which indicate a structural instability with the atoms displacing according to the eigenvector of the mode. Cubic CaMnO$_3$ shows strong instabilities at both the M and R points corresponding to in-phase and out-of-phase octahedral rotations, respectively. SrMnO$_3$ only shows a weak instability at the R point, reflecting the small deviation from ideal cubic in its ground-state structure. Cubic BaMnO$_3$ shows no tilt or rotational instabilities at the M and R points, however a ferroelectric Gamma point mode is unstable. In addition, there also exist a weaker instability at the X point, as previously reported in Ref. 30, which corresponds to a breathing of the oxygen towards and away from the Mn; we do not discuss this mode further. We find that both CaMnO$_3$ and SrMnO$_3$ have G-type antiferromagnetic (AFM) ordering as their magnetic ground state, whereas for BaMnO$_3$ the ferromagnetic (FM) state is the lowest in energy.

Evolution of Structure with Strain

We show the calculated evolution of the structures with strain in Fig. 4a, indicating the evolution of the octahedral rotations and tilts (as defined in Fig. 4b) for CaMnO$_3$ and SrMnO$_3$ and of the ferroelectric polarization for BaMnO$_3$. For comparison purposes all results shown are for G-type AFM ordered structures. Changes in symmetry are indicated by vertical dashed lines with the corresponding tilt patterns labeled, and the regions of phase space in which the materials become polar shaded in blue; the polarization directions are indicated with open blue (for in-plane polarization) and solid blue (out-of-plane) circles. For both CaMnO$_3$ and SrMnO$_3$ the rotation angles are reduced upon tensile strain, whereas tilts are enhanced; the opposite behavior is found for compressive strain, as expected from simple geometric arguments and seen in earlier work[3, 18]. The robust ($a^-a^-c^+$) tilt pattern is maintained throughout the whole strain regime in CaMnO$_3$, except for inaccessibly high (6%) compressive strains where tilts are completely suppressed, whereas the smaller initial rotational angles in SrMnO$_3$ lead to modifications in tilt-pattern with strain; under tensile strain the rotation magnitude becomes zero giving an ($a^-a^-c^0$) tilt pattern, whereas for compressive strain tilts vanish, giving the ($a^0a^0c^-$) tilt pattern. The oxygen octahedra in BaMnO$_3$ remain untilted and unrotated throughout the entire strain range.

Consistent with earlier calculations[8, 19, 43], we find that tensile strain induces a polar distortion in both CaMnO$_3$ and SrMnO$_3$. The distortion is dominated by the Mn ion displacing from the center of its oxygen octahedron and the displacement becomes increasingly pronounced with increasing tensile strain leading to an increase in ferroelectric polarization (Fig. 4a). For CaMnO$_3$, our critical strain value (5.5%) is larger than the previously calculated values of 2.1%[43] and 3.2%[19], which might explain why only precursor signatures of ferroelectricity have been found experimentally[20]. These differences point out the sensitivity of the polar transition to calculation details, such as exchange-correlation functionals (the LDA and the Wu-Cohen GGA[44] were used in Refs. 19 and 43, respectively). We also find a polar state for SrMnO$_3$ under compressive strain, in accordance with a previous theoretical report[8], but the strain range studied is insufficient to induce polarization under compression in CaMnO$_3$. For BaMnO$_3$, the [110] polar displacement of the Mn ion found in the bulk *Amm*2 G-type AFM ground state is enhanced by tensile strain, with compressive strain inducing a *P4mm* structure with polar distortions of the Mn ion along the crystallographic [100] direction.

The evolution of the structures with strain is reflected also in the strain-dependence of the phonon frequencies in the high-symmetry tetragonal reference structures (Fig. 5), again shown for G-type AFM ordering for each material at all strain values.

As we saw above, both R and M-point antiferrodistortive (AFD) modes are unstable in bulk CaMnO$_3$, consistent with the observed *Pbnm* ground state structure, and, as pointed out previously[19, 43] these modes are largely insensitive to strain. The polar modes are stable in the unstrained structure and remain so in the range of compressive strain shown (the out-of-plane mode becomes unstable at the unphysically large compressive strain of 7%). The in-plane polar mode becomes unstable at around 2.5% tensile strain, although competition with the stronger AFD instabilities suppresses the polar ground state until larger strain values.

In the case of SrMnO$_3$, we noted above that the larger Sr$^{2+}$ ions suppress the octahedral rotations relative to those of CaMnO$_3$, and this is reflected in weaker instabilities in the ground state, and stronger sensitivity to strain: Tensile epitaxial strain greater than ~2.5% suppresses all rotational instabilities, and tilt instabilities vanish for compressive strains exceeding ~2%, consistent with the changes in symmetry reported above. As in CaMnO$_3$, the polar modes are characterized by displacement of Mn from the center of its oxygen octahedron and are stable in the bulk structure. In- and out-of-plane polar modes soften with tensile and compressive strain, respectively, with the in-plane mode becoming unstable at ~2.5% tensile epitaxial strain and the out-of-plane for compressive strains larger than ~5%.

In BaMnO$_3$, the large Ba$^{2+}$ cation size stabilizes the tilt and rotational modes over the entire strain range, consistent with its large tolerance factor of 1.097[45]. In contrast, polar instabilities are present for all considered strain values. In the unstrained case, the in-plane and out-of-plane polar modes are degenerate. Upon tensile biaxial strain, two degenerate in-plane polar modes in the [100] direction are increasingly destabilized with strain. Under compressive strain, a polar instability along the [001] direction is enhanced as expected.

Magnetic Order

Finally for this section, we show our calculated total energies for ferromagnetic (FM) and G-type antiferromagnetic (AFM) orderings of the Mn spin moments as a function of strain in Fig. 6. Interestingly, and again consistent with the literature[8], FM ordering is lower in energy than AFM in the polar phase for both $CaMnO_3$ and $SrMnO_3$. (Note that our calculations do not reproduce the intermediate transitions from G-type to C-type to A-type AFM to finally FM ordering obtained for $SrMnO_3$ in Ref. 8 (using a GGA+U functional) likely due to small differences in unit cell volumes used in the two works). We find that $BaMnO_3$, which is always polar, has a FM ground state over the entire strain range. The two minima in the total energy curve for FM $BaMnO_3$ correspond to the energy minima in the *P4mm* (which occurs under compressive strain defined relative to the G-type AFM lattice constant) and *Amm*2 phases; the area within the vertical dashed lines indicates the crossover region where the polarization is oriented intermediate between the [001] and [110] directions. Note that the DFT gap is zero in all FM cases; whether this is an indicator of the onset of metallicity or a consequence of the usual underestimate of the DFT gap will be the subject of future work.

Our observed coupling between the onset of polarization and ferromagnetic order can be rationalized using simple geometric considerations based on the Goodenough-Kanamori rules[46]: For $d^3$ $Mn^{4+}$ cations in octahedral coordination, G-type antiferromagnetic super-exchange is favored by 180° transition metal – oxygen – transition metal bond angles, and is weakened as the M-O-M angle deviates from this angle. At 90° M-O-M bond angles, however, the superexchange is ferromagnetic. The displacement of the Mn ion from the center of its oxygen octahedron in the ferroelectric phase leads to a substantial reduction of the Mn-O-Mn angle and a crossover from predominantly AFM to FM interactions. Such a coupling between magnetic ordering and ferroelectricity was also observed experimentally in $Sr_{0.56}Ba_{0.44}MnO_3$[47], where a significantly reduced polarization was found at the phase transition from the paramagnetic to the G-AFM phase upon cooling. This was attributed to the tendency for 180° Mn-O-Mn bond angles to be favored in the AFM phase, in competition with the polar distortions[48].

To summarize this section, the structural trends in the $CaMnO_3$, $SrMnO_3$, $BaMnO_3$ series with both *A*-site cation and strain that we collected here are exactly as expected based on the increasing *A*-site cation size down the series. While $CaMnO_3$, with its small A-site radius (1.34 Å) and tolerance factor of $0.977^{45}$ has strong octahedral rotations and tilts and needs large strain values to induce ferroelectricity, $BaMnO_3$, with its large B site (1.61 Å) and tolerance factor is always ferroelectric and never shows octahedral rotations or tilts. $SrMnO_3$ is intermediate in its *A*-site size (1.44 Å), tolerance factor of 1.033 $^{45}$ and behavior, with small and strain-dependent tilts, and a ferroelectric ground state at reasonable strain values. The magnetism reflects the presence or absence of ferroelectricity, with the non-polar compounds having the G-type AFM ordering expected for 180º superexchange between $d^3$ $Mn^{4+}$ ions. In the polar phases, the FM state is lower in energy, as a result of the strong deviation of the Mn-O-Mn bonds from 180º and a cross-over to ferromagnetic superexchange.

**Table 1** Calculated pseudo-cubic lattice parameters for CaMnO$_3$, SrMnO$_3$ and BaMnO$_3$ in their fully relaxed perovskite structures, compared to experiment (where available) and earlier calculations using the local spin density (LDA) **[30]** and PBE-sol **[18]** functionals.

|  |  |  | a (Å) | b (Å) | c (Å) |
|---|---|---|---|---|---|
| CaMnO$_3$ | Measured **[22]** |  | 3.727 | 3.727 | 3.727 |
|  | Calculated, PBEsol **[18]** |  | 3.725 | 3.725 | 3.715 |
|  | Calculated, this work |  | 3.729 | 3.729 | 3.722 |
| SrMnO$_3$ | Measured **[49]** |  | 3.805 | 3.805 | 3.805 |
|  | Calculated, this work | *Pbnm* | 3.799 | 3.799 | 3.790 |
|  |  | *R*-3c | 3.796 | 3.796 | 3.796 |
| BaMnO$_3$ | Calculated, LSDA **[30]** | *Amm*2 | 3.84 | 3.84 | 3.84 |
|  | Calculated, this work | *Amm*2 (G-AFM) | 3.922 | 3.922 | 3.895 |
|  |  | *P4mm* (FM) | 3.870 | 3.870 | 4.133 |

**Oxygen vacancy stability and ordering**

Next we study how the interplay between strain, structure, magnetism and A-site cation described above further couples and/or competes with the formation and stability of oxygen vacancies. On formation of a neutral oxygen vacancy, we expect the two electrons from the formally $O^{2-}$ ion to transfer to the adjacent Mn ions reducing them from formally $Mn^{4+}$ to $Mn^{3+}$, which is expressed in Kröger Vink notation as:

$$2Mn_{Mn}^{x} + O_{O}^{x} \rightarrow 2Mn'_{Mn} + V^{\cdot\cdot}_{O} + \frac{1}{2}O_2 \quad .$$

In this notation each species is assigned a charge relative to its formal charge in the stoichiometric compound, with the superscripts "x", "¨" and "`" denoting neutral, positive (2+) and negative (-1) charges, respectively. The subscripts indicate the identity of the lattice site. In the localized electron limit the reduction of manganese is accompanied by an increase in ionic radius from 0.53Å for 6-coordinated $Mn^{4+}$ to 0.645 Å for $Mn^{3+}$, and therefore we expect oxygen vacancy formation to be favored when the lattice is expanded by A-site cations or by tensile strain.

**Table 2** Calculated formation energies of oxygen vacancies in the ground-state bulk structures. There are two inequivalent lattice sites which we label IP and OP, in anticipation of the strain dependence that we study in the next section.

|  | IP $V_O$ (eV) | OP $V_O$ (eV) |
|---|---|---|
| $CaMnO_3$ (*Pbnm*) | 1.86 | 1.87 |
| $SrMnO_3$ (*Pbnm*/*R-3c*) | 1.27 | 1.24 |
| $BaMnO_3$ (*P4mm*) | 0.81 | 0.51 |

In Table 2 we show our calculated oxygen vacancy formation energies for all three compounds in their bulk structural and magnetic ground states. The oxygen chemical potential $\mu_0$ was set to -5eV, corresponding to typical growth conditions under air[18]. We see that, as expected, the formation energy is largest for $CaMnO_3$ and smallest for $BaMnO_3$, with $SrMnO_3$ intermediate. Interestingly, we find that for $CaMnO_3$ and $SrMnO_3$, which are antiferromagnetic in their ground states, the lowest energy state with one oxygen vacancy is in fact ferromagnetic. This is in part because the smaller band gaps in the FM states more readily accommodate additional electrons leading to lower electronic energy costs to forming vacancies in the FM structures. Closer inspection of the electronic structure reveals in addition that the charge compensating electrons left by the oxygen vacancies are delocalized over all Mn ions in the FM case, whereas if the ordering is constrained to G-type AFM they localize on the two Mn ions adjacent to the vacancy. This implies that an additional stabilization mechanism for ferromagnetism is the double exchange mechanism in which carriers can lower their kinetic energy through delocalization provided that the magnetic moments on all Mn ions have the same orientation. We emphasize that our finding is for a particular supercell size and oxygen vacancy concentration, and that FM is less likely in the dilute limit for which the oxygen vacancy formation energy is formally defined, and also that the quantitative details are of course sensitive to the exact magnitude of the band gaps in the different magnetic configurations. $BaMnO_3$ is already FM in its ground state and does not change its magnetic ordering when an oxygen vacancy is formed.

Next, we investigate the strain dependence of the oxygen formation energies. We begin by revisiting $CaMnO_3$, for which the strain dependence of the oxygen vacancy formation energies was already reported in the case where the magnetism is constrained to be G-type AFM in the presence of the vacancy[18]. Our results are shown in Fig. 7. for the two symmetry-inequivalent oxygen vacancy positions shown in Fig. 1, one in the Mn-O-Mn bonds within the strained plane (IP) and one along the out-of-plane lattice vector (OP). We note that our trends with strain are similar to those of Ref. 18, with our formation energies slightly lower when unstrained (about 0.2eV) and with a more pronounced lowering with tensile strain, caused by the additional stabilization when allowing FM ordering upon vacancy formation. (When constraining the magnetic ordering to G-AFM, we obtained formation energies in full accordance with Ref. 18). As reported previously, with increasing tensile strain (and correspondingly increasing volume), the formation energies of IP and OP vacancies both decrease, with the IP vacancies becoming most strongly favored suggesting a route to controlling oxygen vacancy ordering. Intriguingly, the decreasing trend in vacancy formation energy with tensile strain is either reduced or reversed (depending on the vacancy type) upon entering the polar region. Formation energies for the non-polar structures are inserted as gray lines in Fig. 7, to emphasize the effect of the polar distortions, even though not representing the true ground state in the polar regime.

The vacancy formation energy is approximately constant under compressive strain. This was rationalized in Ref. 18 as a competition between the chemical expansion effect and shifts in the energy levels caused by changes in the crystal field.

The interplay between the formation of the polar state and the vacancy formation energy can be seen more clearly in the case of SrMnO$_3$ under tensile strain, which shows the same general trends as CaMnO$_3$ but in which the polar region extends over a greater strain range (Fig. 6 middle panel). We rationalize this behavior as follows: it is established that the stabilization of polar distortions in the $d^3$ perovskite manganites arises from the charge transfer from filled oxygen 2p to empty manganese 3d $e_g$ orbitals[30, 50] by analogy to the hybridization with the empty Ti 3d $t_{2g}$ orbitals in the prototypical $d^0$ perovskite oxides such as BaTiO$_3$[51]. When an oxygen vacancy is formed, the two charge compensating electrons occupy the manganese $e_g$ orbitals [18], inhibiting the formation of partial covalent bonds between Mn 3d $e_g$ and O 2p. The electronic structure changes associated with oxygen vacancy formation thus effectively suppress the microscopic mechanism for stabilizing ferroelectric dipoles. A consequence of this competition between oxygen vacancy formation and ferroelectric polarization, combined with their mutual coupling to strain, is that a system can respond to tensile strain either by developing a polar distortion, or by introducing oxygen vacancies. The presence of oxygen vacancies, therefore will suppress the formation of the ferroelectric state by providing an alternative pathway for accommodating the tensile strain; conversely once a polar phase is formed the formation of oxygen vacancies is more energetically costly. Indeed in BaMnO$_3$ (lower panel), which is polar throughout the entire strain range, the strain-dependence of the oxygen vacancy formation energy is exactly opposite to that expected by usual chemical expansion arguments, with both compressive and tensile strain increasing the formation energy consistent with their increasing the polarization. Since none of the strain is accommodated by oxygen rotations or tilts, we find also a strong anisotropy between the behavior of the IP and OP vacancies, with the former having lower energy at tensile and higher at compressive strain and the opposite behavior for the latter. This can be rationalized by the local change in volume at the vacancy site on strain – under tensile strain the IP site is increased in volume and the OP decreased and vice versa –and is more significant in the absence of compensating rotations or tilts.

We see that the oxygen vacancy formation energies in the II-IV manganites are the result of a complex interplay between strain, A-cation size, magnetic ordering (G-AFM or FM), polar distortions and antiferrodistortive rotations and tilts, and while simple chemical expansion guidelines are often helpful, the full picture is clearly more complicated. In particular oxygen vacancy formation energies are increased by the presence of polar distortions and decreased by ferromagnetic ordering; conversely their presence suppresses ferroelectricity and favors ferromagnetism.

The concentration and ordering of oxygen vacancies can be controlled experimentally through careful adjustment of the thermal and atmospheric history of the samples, including both growth conditions and post-growth annealing. While it has been known for decades that oxygen vacancies influence ferroic order parameters, the full extent of these interactions, especially on an atomic scale, remains unknown for the vast majority of systems. Strong coupling, or competition, between ferroic order parameters and oxygen vacancies suggests that *in situ* annealing under electric or magnetic fields could modulate the oxygen vacancy formation energy and especially ordering. This could open a route to control oxygen vacancy populations, beyond the global concentration, by engineering concentration gradients across thin films.

We summarize the coupled and competing mechanisms identified in the $AMnO_3$ (A=Ca,Sr,Ba) system schematically in Fig. 8. The rich spectrum of structural, magnetic and electronic properties points towards the active use of oxygen vacancies as a parameter in design of functional material properties. We believe that our findings apply not only to the II-IV manganites investigated here, but more broadly to ferroic transition metal oxides and hope that they further motivate the exploration of oxygen vacancies as an additional control parameter in design of functional material properties.

**Acknowledgements**

This work was financially supported by the ETH Zürich (CF and NAS) and by the ERC Advanced Grant program, No. 291151 (CF and NAS). Computational resources were provided by the Swiss National Supercomputing Centre (CSCS) under project ID mr6 and the Norwegian Metacentre for Computational Science, Sigma2, through the project nn9264k.

# Figures

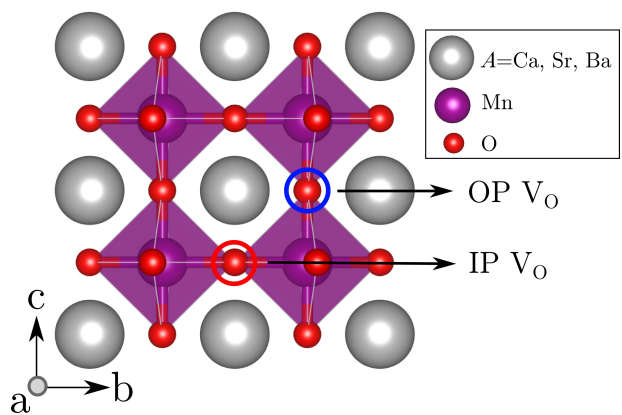

**Figure 1** Inequivalent oxygen vacancy sites, one breaking an Mn-O-Mn bond within the biaxial strain-plane (IP Vo) and one breaking an Mn-O-Mn bond perpendicular to the strain plane.

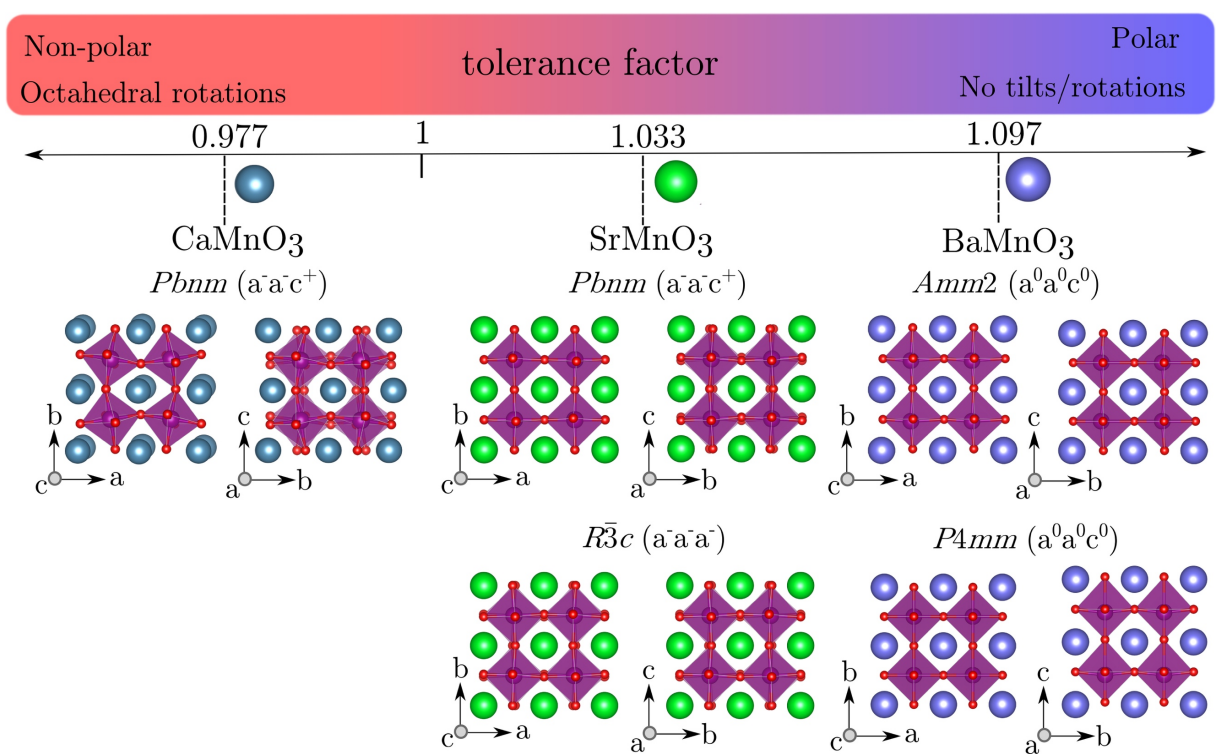

Figure 2 Calculated structures of $CaMnO_3$, $SrMnO_3$ and $BaMnO_3$ structures, with their A-site cation sizes and tolerance factors indicated. The *Pbnm* and *R-3c* structures calculated for $SrMnO_3$ are indistinguishable in energy within the accuracy of our calculations; likewise the likewise the two different ferroelectric polarization directions in BaMnO3, *Amm*2 (having AFM magnetic order) and *P4mm* (having FM magnetic order) are very close in energy.

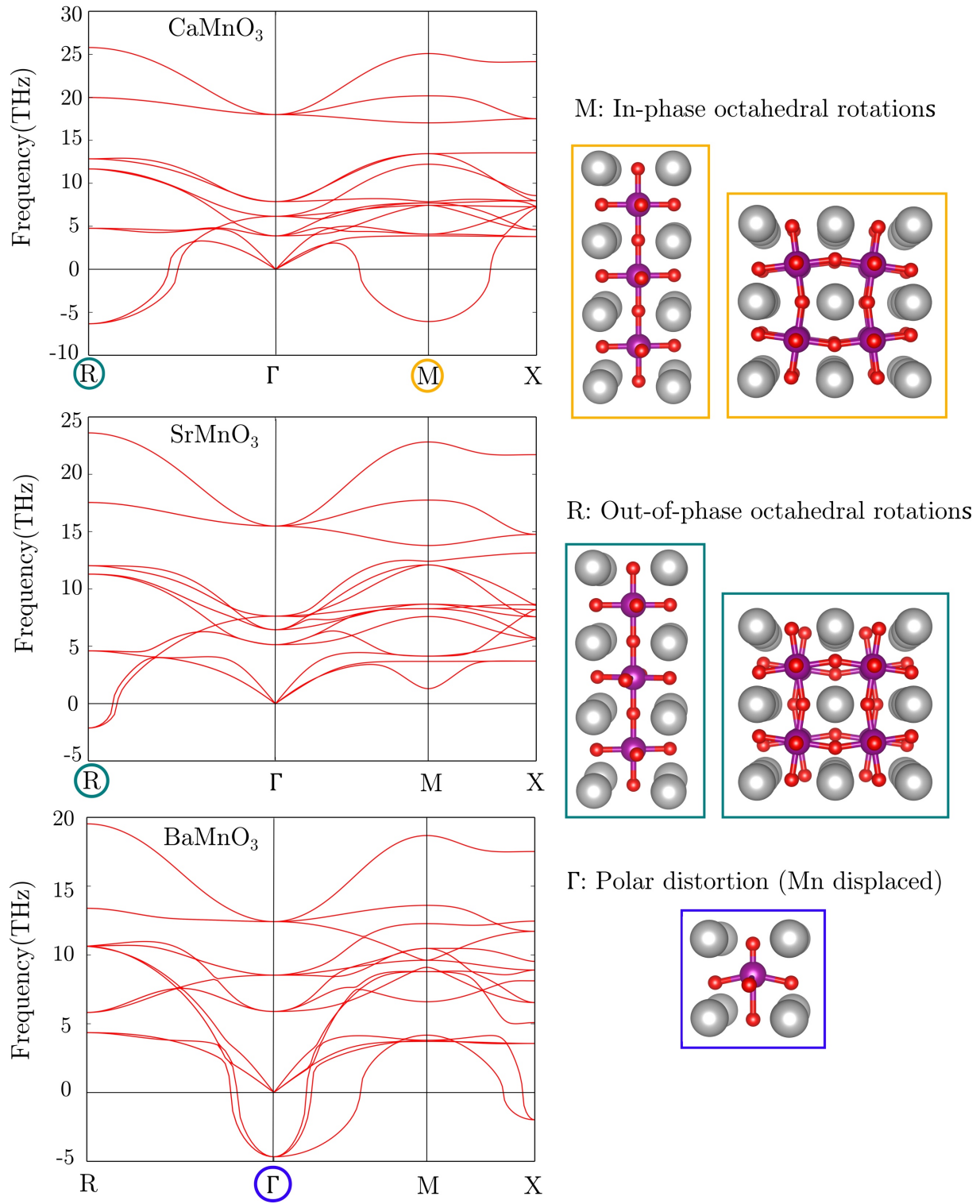

**Figure 3** Calculated phonon dispersions for bulk G-AFM CaMnO$_3$ (top), SrMnO$_3$ (middle) and BaMnO$_3$ (lower) in the high symmetry *Pm*-3*m* structure. Unstable imaginary frequencies are plotted as negative values. Schematics of the unstable rotational and polar modes are shown on the right.

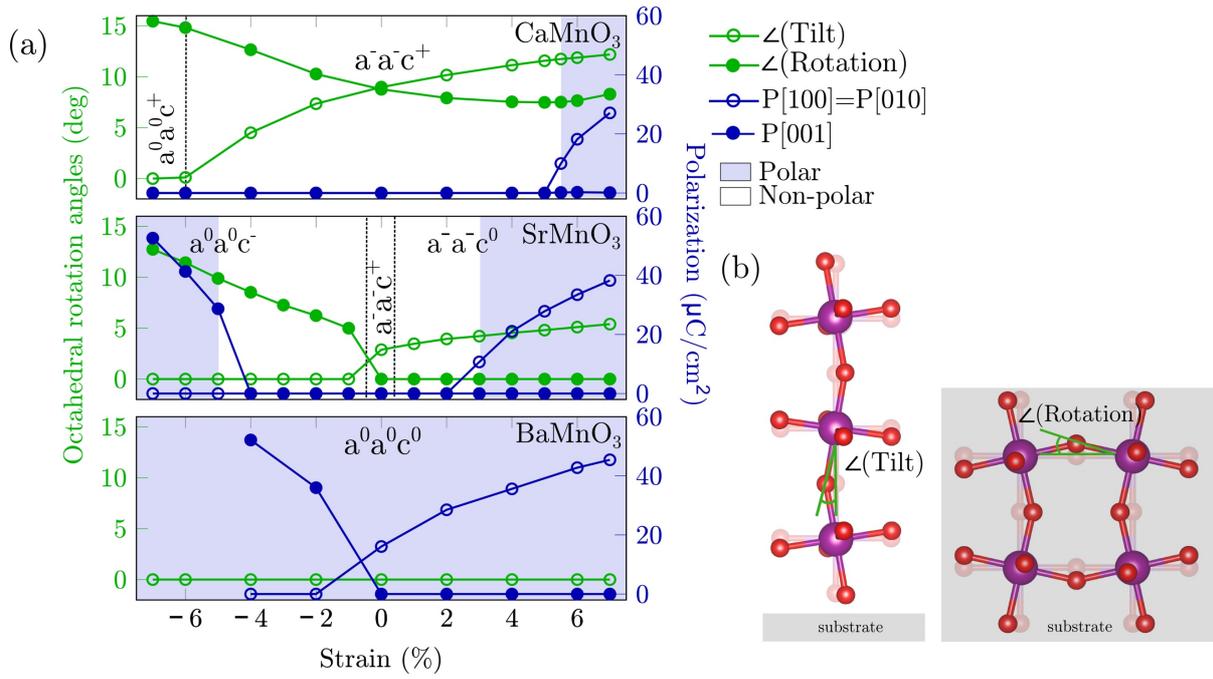

**Figure 4** (a) Evolution of octahedral rotation angles (defined as in (b)) and polarization with biaxial strain for CaMnO3 (upper panel), SrMnO3 (middle panel) and BaMnO3 (lower panel). All calculations are for G-type AFM ordering. Changes in AFD pattern are indicated by vertical dashed lines and polar regions are shaded in blue. Note that AFM $BaMnO_3$ becomes metallic at compressive strains larger than 4% and so the Berry phase polarization is not defined at these values.

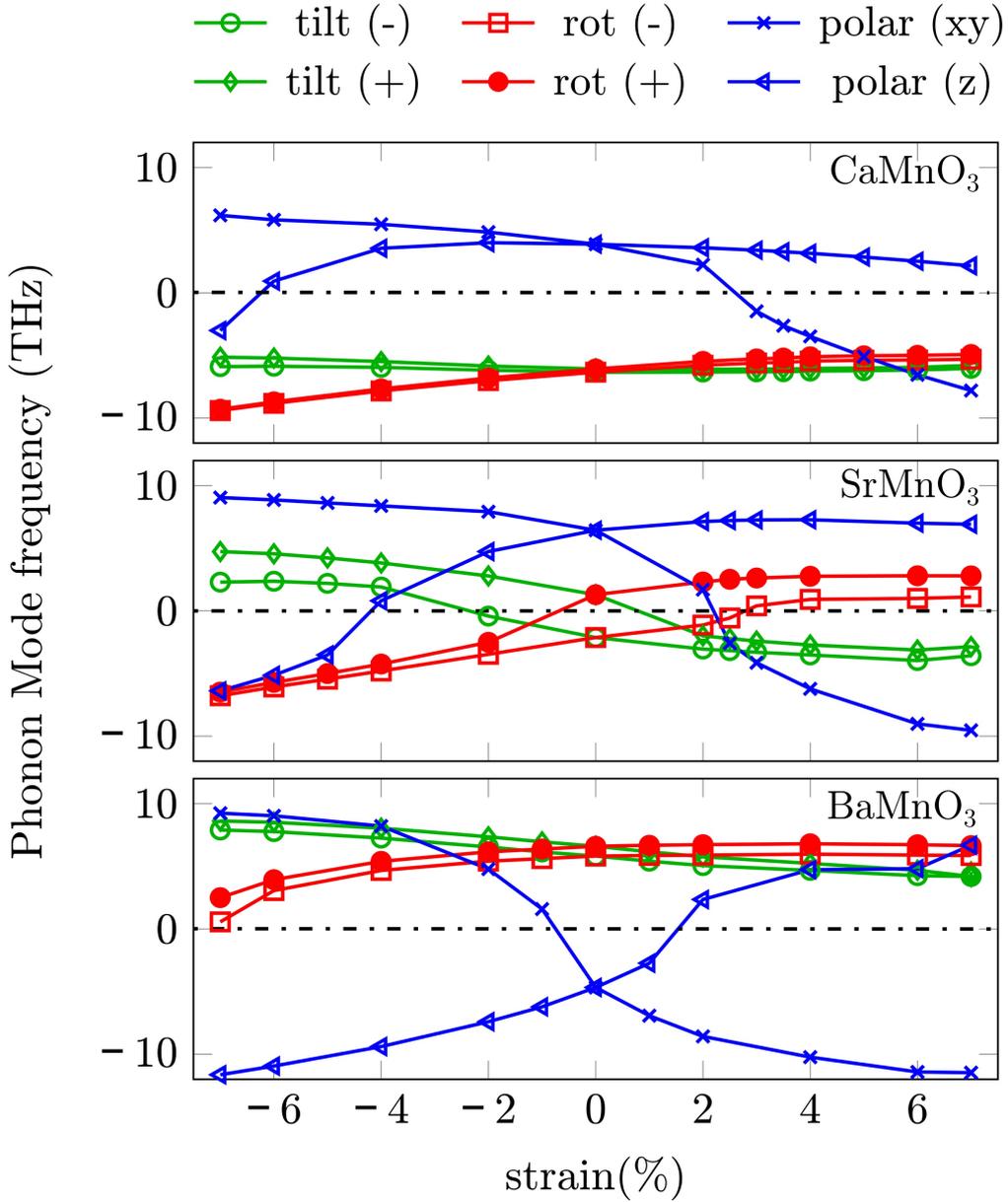

**Figure 5** Evolution of phonon frequencies as function of (100) biaxial strain in the ideal (Pmmm) perovskite structures for CaMnO$_3$, SrMnO$_3$ and BaMnO$_3$ (G-AFM ordering). Phonon modes are grouped into polar modes (blue), antiferrodistortive rotational modes (red) and tilt modes (green). In-phase tilts/rotations are indicated by (+) whereas out-of-phase tilts/rotations are indicated by (-). Upon increasing the size of the A cation from Ca to Sr to Ba, tilts/rotations are increasingly suppressed, whereas polar distortions are increasingly enhanced. Tilts and rotational modes are less affected by epitaxial strain than polar modes. Tensile strain enhances in-plane polarization and suppresses out-of-plane polarization, and vice versa for compressive strain.

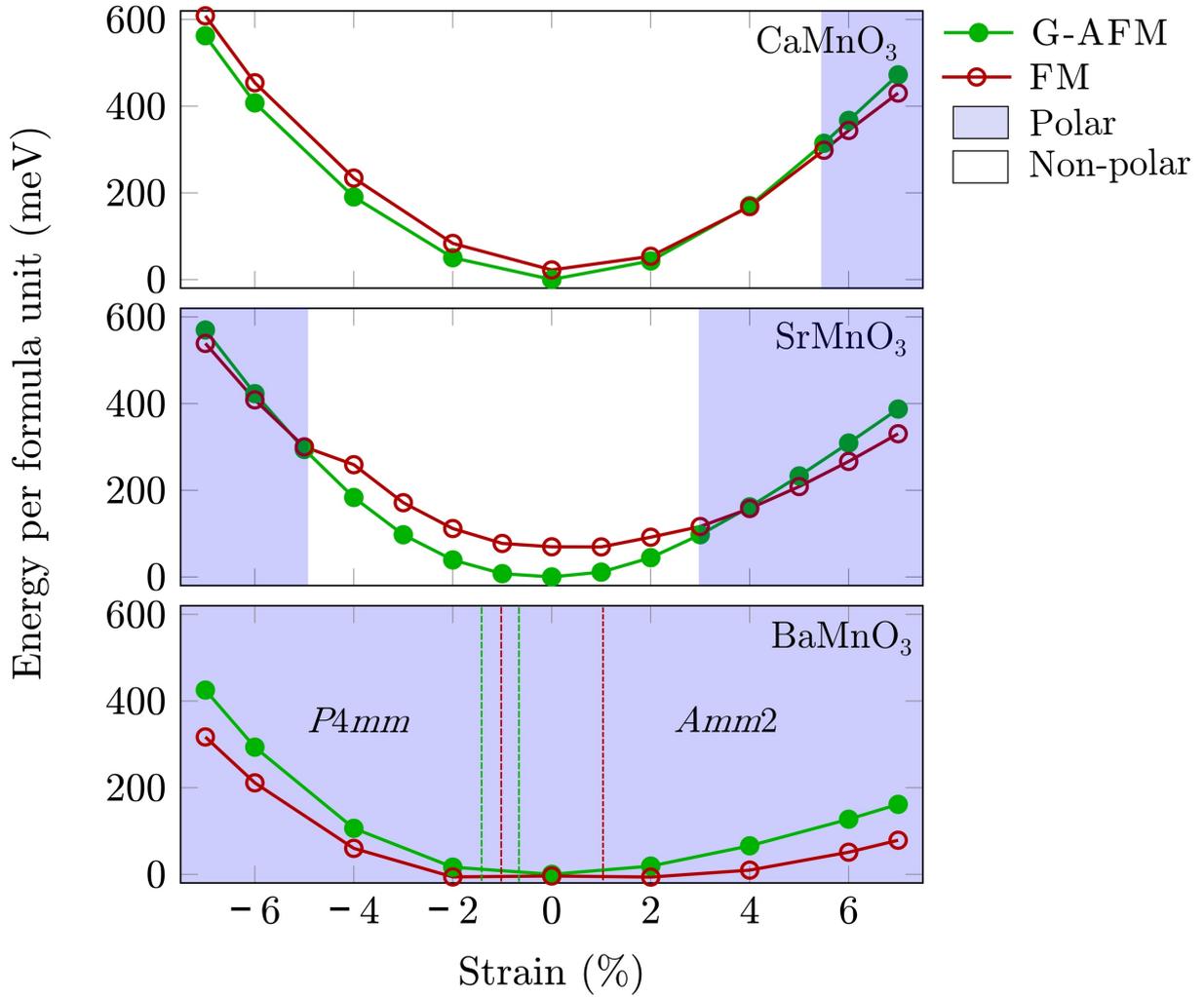

**Figure 6** Energy per formula unit as function of biaxial strain for G-AFM and FM orderings of the Mn moments. Energies are given relative to the unstrained G-AFM total energy for each material system, and polar regions are shaded in blue. Note that FM ordering is always lower in energy when the structure is polar; in these regions the DFT gap closes to zero. Vertical lines on the BaMnO$_3$ figure indicate the range with $P1$ symmetry within which the polarization reorients between [110] and [001] directions.

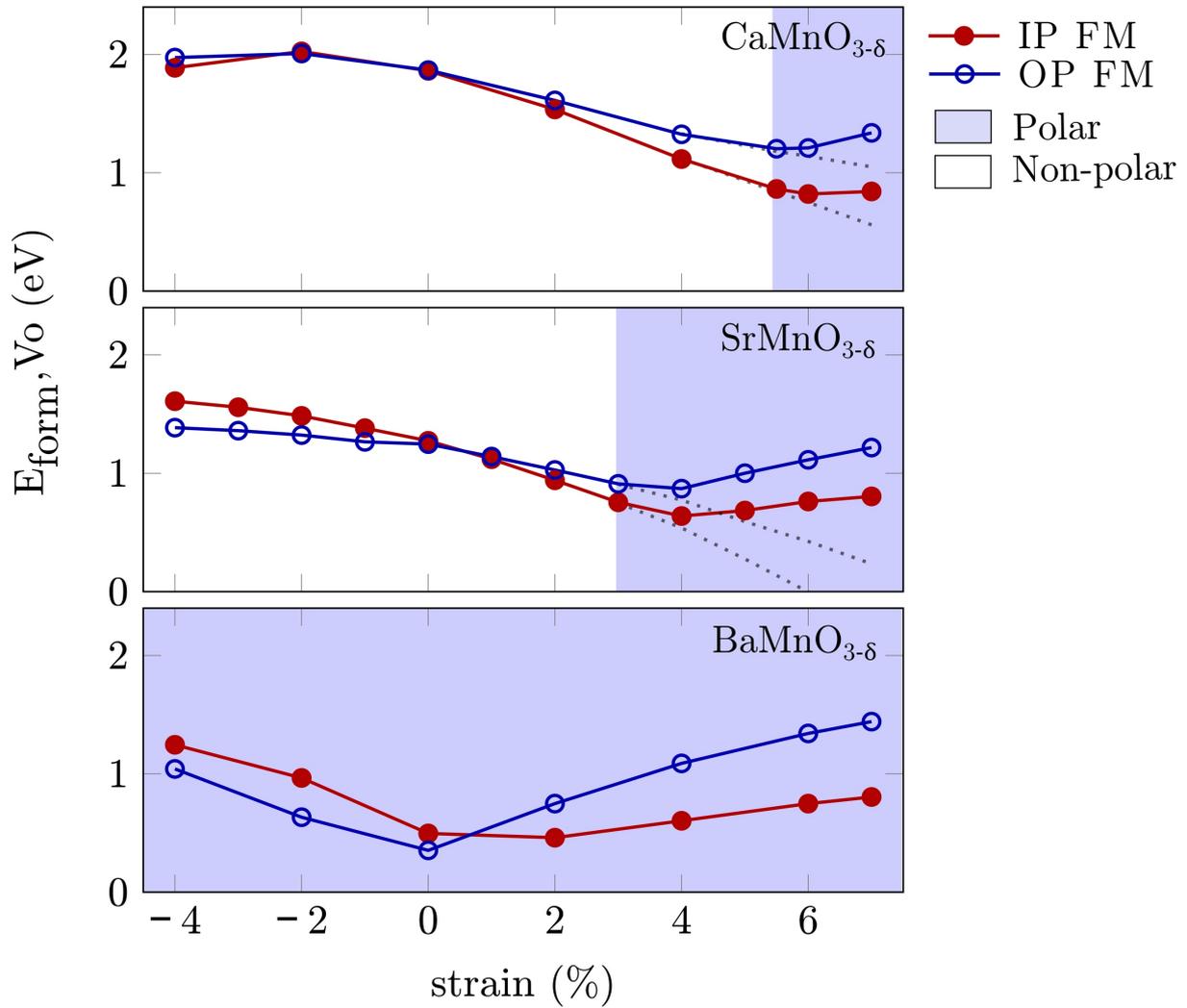

**Figure 7** In-plane (IP) and out-of-plane (OP) oxygen vacancy formation energies for $CaMnO_3$, $SrMnO_3$ and $BaMnO_3$. As a guide for the eye, formation energies for the non-polar structures are inserted as gray lines, even though not representing the true ground state in the polar regime.

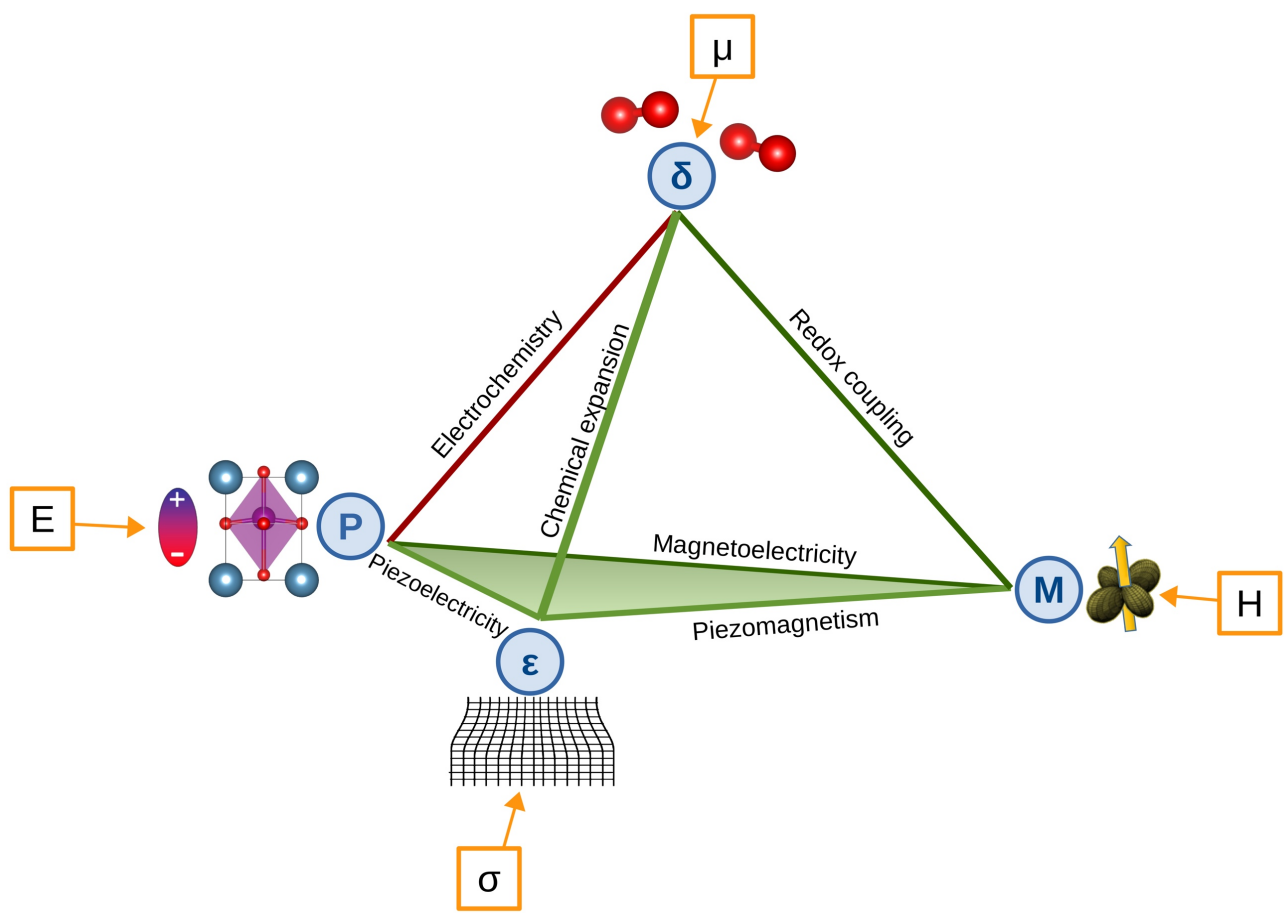

**Figure 8** Schematic illustration of coupling and competing mechanism as discussed in this paper. Green lines along the edges indicate cooperative behavior whereas red lines denote a competition.